\documentclass[sigconf,nonacm]{acmart}
\renewcommand\footnotetextcopyrightpermission[1]{}
\usepackage{booktabs}
\usepackage{tabularx}
\AtBeginEnvironment{thebibliography}{\interlinepenalty=10000}
\begin{document}
\title{Faithful Query Resolution for Conversational Search}
\subtitle{UiS\&UP at iKAT SCAI 2026}
\author{Saber Zerhoudi}
\affiliation{\institution{University of Passau}\city{Passau}\country{Germany}}
\email{szerhoudi@acm.org}
\author{Ivica Kostric}
\affiliation{\institution{University of Stavanger}\city{Stavanger}\country{Norway}}
\email{ivica.kostric@uis.no}
\author{Jelena Mitrovi\'{c}}
\affiliation{\institution{University of Passau}\city{Passau}\country{Germany}}
\email{jelena.mitrovic@uni-passau.de}
\renewcommand{\shortauthors}{Zerhoudi et al.}
\begin{abstract}
Conversational query rewriting makes context-dependent requests explicit for retrieval, but additional context can introduce an information need the user did not express. This raises a question about the interaction between query fidelity and retrieval breadth. This paper examines that interaction in behavior-adaptive conversational search, using iKAT SCAI 2026 as the evaluation setting. We compare five query-generation configurations within a behavior-adaptive assistant using BM25 and SPLADE retrieval. A minimal-faithful resolver preserves the user's wording and resolves only context-dependent references; weighted reciprocal rank fusion combines its rankings with those from a learned rewriter and the raw utterance. Offline evaluation shows that adding these formulations recovers recall over the faithful query alone, with a slight reduction in nDCG@10. Across our five iKAT SCAI 2026 submissions, the weighted configuration obtains the highest mean retrieval scores, reaching nDCG@5 of 0.534, a 5.3\% relative improvement over the strongest alternative. However, this retrieval advantage does not coincide with the highest conversational satisfaction scores. These findings motivate separating faithful query interpretation from retrieval expansion, while assessing their effects on \mbox{conversational utility independently.}
\end{abstract}
\maketitle
\sloppy
\raggedbottom
\finalhyphendemerits=100000

\section{Introduction}
Conversational search systems must interpret an utterance in the context of earlier turns. Resolving a pronoun can make an information need explicit; adding plausible topics or personal facts can change it. For example, the correction ``I mean black garlic.'' already names its subject. Rewriting it as ``how to use black garlic in gourmet cooking'' introduces a use case that the user has not specified. This distinction motivates a simple question: \emph{How much context should a resolver add, and how should \mbox{alternative queries contribute to retrieval?}}

The Interactive Knowledge Assistance Track (iKAT) combines conversational retrieval with a personal textual knowledge base (PTKB)~\citep{Aliannejadi:2023:TREC}. In iKAT SCAI 2026, the assistant interacts with simulated users through Sim.API~\citep{Gohsen:2026:SIGIR}, retrieves ClueWeb22 passages, and returns responses, citations, and PTKB provenance. A shared controller supports five submitted configurations that differ primarily in query generation. This setting lets us examine query resolution inside a working conversational system, alongside \mbox{the limits of its response behavior.}

Our study separates two decisions. \textbf{Resolving references} makes the current request interpretable while preserving its wording. \textbf{Broadening retrieval} adds independently generated queries and controls their influence through fusion weights. Our offline experiments favor faithful resolution over the earlier combined resolver on nDCG@10, but additional queries recover recall. The weighted configuration also obtains the highest official mean retrieval scores among our submissions. These observations support a design choice within this system; they do not establish that minimal rewriting is optimal across conversational search tasks.

We make two contributions. \textbf{(1) Query design:} we describe minimal-faithful resolution and its integration with weighted hybrid retrieval, and report the observed ranking--recall tradeoff. \textbf{(2) Evaluation:} we compare five interactive submissions and distinguish retrieval gains from response quality, explicitly separating offline evaluation scores from official LLM judgments. 

\section{Background}
Conversational retrieval uses earlier turns to interpret the current request, as in the TREC Conversational Assistance Track~\citep{Dalton:2020:arXiv}. iKAT adds user-specific knowledge through a PTKB~\citep{Aliannejadi:2023:TREC}, making personal context available alongside the conversation. PersonaRAG~\citep{Zerhoudi:2024:SIGIR-ws} uses user-centric agents to adapt retrieval and generation to user data and interactions. These settings motivate the query-design question we examine: which contextual information is needed to resolve the current utterance, and which additions broaden the search beyond \mbox{what the user expressed?}

Preserving the request still leaves a retrieval problem: the user's wording may differ from that of relevant passages. Our system combines BM25~\citep{Robertson:2009:FnTIR} and SPLADE~\citep{Formal:2021:arXiv} rankings using reciprocal rank fusion (RRF)~\citep{Cormack:2009:SIGIR}, which does not require comparable retrieval scores. Query rewriting can supply alternative formulations, each producing a ranking that contributes to the fusion. We examine how these alternatives complement a faithful query and how their relative weights affect \mbox{the retrieved evidence.}

The effect of that evidence also depends on the ensuing interaction. User simulation supports evaluation of how an assistant answers, clarifies, and responds to feedback. UXSim~\citep{Zerhoudi:2025:CIKM} combines structured simulation components with an adaptive LLM agent, separating behavioral constraints from language generation on the user side. In our iKAT evaluation, Sim.API~\citep{Gohsen:2026:SIGIR} supplies the simulated users, while our behavior detector adapts the assistant to their observed interaction. This setting lets us compare retrieval scores with response assessments across the same submissions, examining whether broader retrieval also improves \mbox{the simulated conversations.}

\section{Query Fidelity and Breadth}
\label{sec:problem}
Query generation has two roles in this system: interpreting the request and retrieving enough evidence to answer it. Our system handles these roles separately. We distinguish three issues that guide the design and determine what \mbox{the evaluation can establish.}

\subsubsection*{Resolving Context Faithfully}
A user can leave an entity unnamed or refer to a previous answer. The resolver must recover that reference before retrieval can match the intended subject. However, additional context can introduce a different task, a preferred use case, or a personal constraint that the user does not mention. A useful query therefore needs enough context to be interpretable while remaining faithful to what was asked. The minimal-faithful resolver restricts its edits to this interpretation step; inspection of query examples motivated the restriction, without establishing \mbox{how often over-elaboration occurs.}

\subsubsection*{Combining Alternative Formulations}
A faithful query can still miss relevant passages because it retains the user's vocabulary. Additional formulations can retrieve complementary evidence. When each ranking receives the same weight in RRF, however, passages repeatedly retrieved by alternative queries accumulate support that can outweigh the faithful query's results. This motivates assigning the faithful query greater weight while preserving alternatives in the candidate pool. The weight determines each query's contribution to the fused ranking. Its effect depends on the quality and overlap of the retrieved lists.

\subsubsection*{Assessing Conversational Utility}
Retrieval scores measure properties of passage rankings, while a useful interaction also depends on how the assistant answers, clarifies, and responds to feedback. Simulated interaction introduces a further distinction between performance under the specified user model and behavior with real users. \citet{Zerhoudi:2022:ECIR} evaluate simulated search sessions against real interaction data, illustrating that simulator realism is itself an evaluation target. We do not assess Sim.API's behavioral fidelity here. The supplied response assessments measure performance in these simulated conversations; human preferences are \mbox{not measured in our study.}

These issues yield two research questions. \textbf{RQ1:} Does faithful resolution improve offline retrieval, and what changes when other queries are added? \textbf{RQ2:} Does the selected configuration improve official retrieval and response scores? The first examines the ranking--recall tradeoff under a shared offline retriever; the second examines whether the selected configuration's retrieval ordering coincides with response-quality gains in interactive runs.

\section{System and Evaluation}
\label{sec:eval}
We address these questions through our offline experiments and five official submissions. The system design separates faithful resolution, alternative query generation, and interaction control; the evaluation keeps offline retrieval and \mbox{official interaction scores distinct.}

\subsection{Experimental Design}
The offline comparison varies the query formulation while retaining the hybrid retriever. The manual query provides a reference for the same retrieval stack, and the faithful-only condition exposes what is lost when alternative formulations are removed. We combine the retrieved rankings using RRF, giving passages retrieved by the faithful query twice the contribution of passages at the same rank retrieved by an alternative query. This comparison changes both the query set and the weight, so it does not isolate \mbox{the effect of weighting alone.}

The official comparison uses independently generated interactive runs. It tests the submitted configurations as deployed, including their query-generation choices and robustness revisions. Official retrieval and response assessments are reported separately because they measure different outcomes. Neither comparison establishes a causal effect on human satisfaction.

\subsection{System}
\label{sec:method}
\begin{figure*}[t]
\centering
\includegraphics[width=0.9\textwidth]{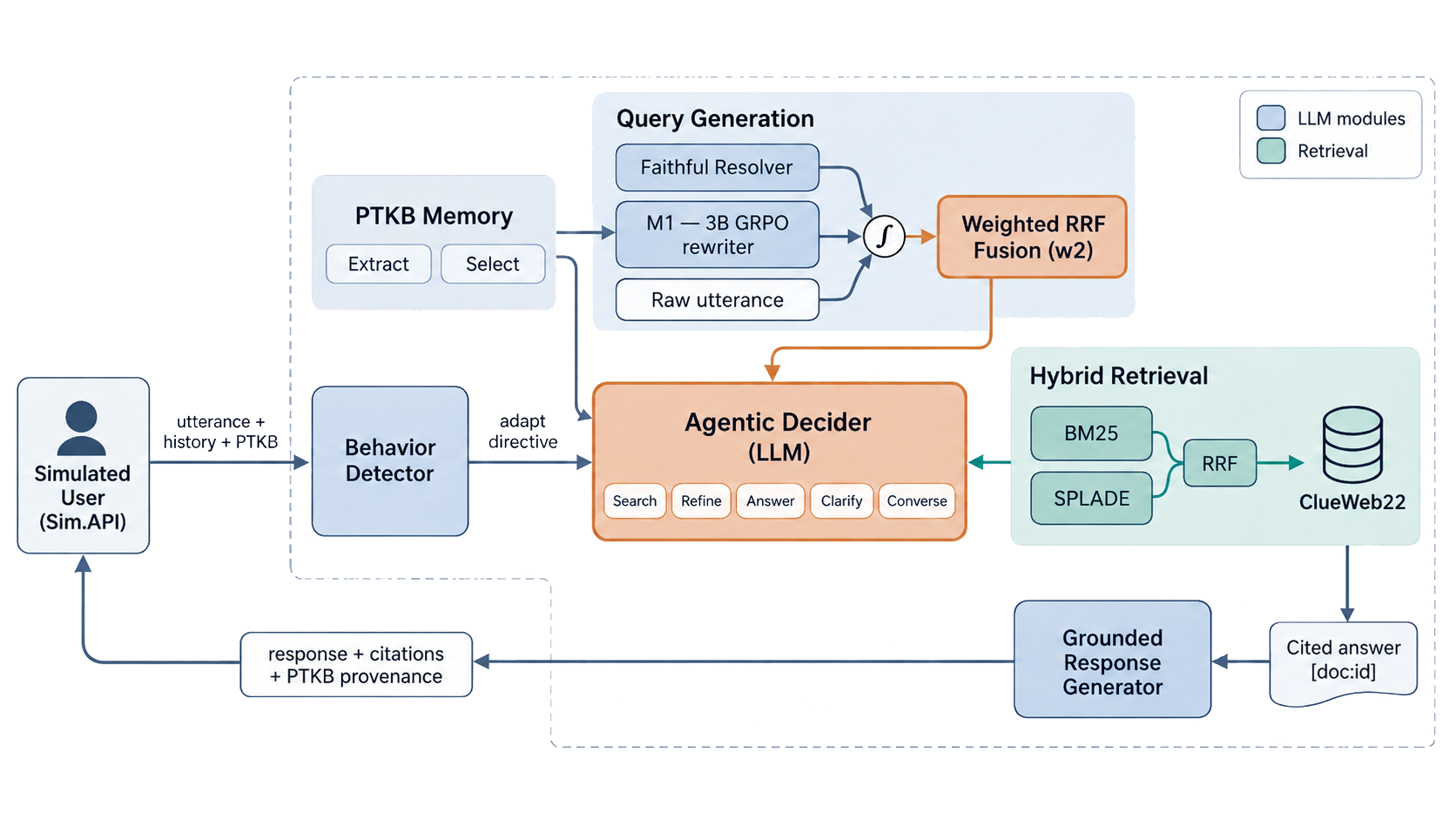}
\caption{System architecture. Sim.API provides an utterance, history, and PTKB. The behavior detector adapts the agentic decider, which requests retrieval and produces a grounded response. Query generation combines the faithful resolver, the 3B GRPO rewriter M1, and the raw utterance. ``w2'' doubles the faithful query's contribution for passages at the same rank. In implementation, weighted RRF aggregates the BM25 and SPLADE lists for each query in one sum; the two fusion boxes depict the query and retriever dimensions of that operation. The grounded response generator is part of the decider. The integral-shaped symbol denotes query combination, and \texttt{[doc:id]} denotes an inline passage citation.}
\Description{Architecture connecting simulated users, PTKB memory, query generation, a behavior detector, an agentic decider, hybrid retrieval, and grounded responses.}
\label{fig:architecture}
\end{figure*}

The system combines a reference-preserving query generator with a behavior-adaptive controller and a hybrid retriever (Figure~\ref{fig:architecture}).

\subsubsection{Preserving the Information Need}
The minimal-faithful resolver receives the current utterance, history, and PTKB. It resolves pronouns and elliptical expressions to their antecedents, retaining the rest of the user's wording. An utterance that is already interpretable remains unchanged. In the black-garlic example, it retains ``I mean black garlic.''; for ``Which of the aforementioned ones is natural, not pill-based?'', it substitutes the previously listed alternatives. These prompt examples illustrate the intended behavior. Resolver accuracy has not been measured through \mbox{independent annotation.}

M1 is our 3B query rewriter, trained with group relative policy optimization (GRPO)~\citep{Shao:2024:arXiv} using retrieval rewards against iKAT-2023 qrels. The query generator places the faithful query first, appends M1 rewrites and the raw utterance, and removes duplicate query strings. Additional formulations broaden the candidate pool without requiring the faithful query itself to include speculative context.

\subsubsection{Weighting the Faithful Query}
For each unique query $q\in Q$, BM25 and SPLADE retrieve ranked passage lists from the ClueWeb22 iKAT index~\citep{Overwijk:2022:SIGIR}. Let $r_{q,m}(d)$ be the one-based rank of passage $d$ for query $q$ and retriever $m$. The fused passage score is
\begin{equation}
 s(d)=\sum_{q\in Q}\sum_{m\in\{\mathrm{BM25},\mathrm{SPLADE}\}}
 \frac{w_q\,\mathbf{1}[d\in L_{q,m}]}{60+r_{q,m}(d)},
\end{equation}
where $L_{q,m}$ is the retrieved list and a missing passage contributes zero. We set $w_q=2$ for the faithful query and $w_q=1$ for each other query, so a passage retrieved by the faithful query contributes twice as much as a passage at the same rank retrieved by an alternative query. This applies to both BM25 and SPLADE rankings. This is a single aggregation over all lists. The faithful query has greater influence, but passages supported by multiple alternative queries can still outrank its results. No \mbox{neural reranker follows fusion.}

\subsubsection{Controlling the Interaction}
A behavior detector summarizes signals such as brief replies, repeated requests, and closing intent. The agentic decider reads this directive and chooses search, refine, answer, clarify, or converse. Each decision uses a structured LLM call; search actions can lead to further decisions within the same user turn until a tool-call budget requires an answer. Thus, one user turn can contain multiple model calls. The deployed prompt favors answering once evidence is available, treats disputes as requests for a revised answer, and discourages consecutive clarification questions. The decider sees at most 20 passage snippets of 300 characters each. Its default generation budget is 900 tokens, with temperature 0.2. The reported comparisons do not establish optimal values \mbox{for these settings.}

The decider selects relevant provided PTKB indices and writes the answer from retrieved evidence. A separate extractor maintains newly stated durable user facts, and a selector identifies those relevant to the turn. Provided and discovered facts are selected separately, then combined for PTKB provenance, so the provided-fact decision remains tied to the supplied statement indices. Inline \texttt{[doc:id]} citations are filtered against submitted passage identifiers, and the response is capped to the API length limit. Identifier consistency alone does not establish that a passage supports the associated claim. The response filter uses a conservative limit of 240 word and punctuation tokens against the API ceiling of 250.

\subsection{Experimental Setup}
\emph{Offline evaluation.} Our experiments use iKAT-2023 passage qrels, preceding reference dialogue turns, and the same hybrid retriever. Offline retrieval uses lists of up to 1,000 passages per query and retriever; the deployed searcher uses 100. We retain the previously recorded nDCG@10 and recall@1,000 aggregates. The metric implementation uses linear relevance gains for nDCG and positive grades for recall. The manually resolved query provides a reference score for the same retrieval stack. We use iKAT-2023 for M1 training, resolver design, and offline evaluation.

\emph{Interactive evaluation.} We compare five submitted runs across four simulated user types (u1--u4). Retrieval and response quality are evaluated using LLM-based assessments. We average retrieval scores equally across user types and report mean dialogue scores over 44 assessments per run. Response quality is also assessed at rubric level, with the number of assessments varying across runs. These comparisons use \mbox{descriptive aggregates without significance tests.}

\begin{table}[t]
\caption{Submitted configurations. Run identifiers use the prefix \texttt{uis-up-run-}. R4 and R5 \mbox{also include later robustness fixes.}}
\label{tab:runs}
\small
\begin{tabularx}{\columnwidth}{@{}lX@{}}
\toprule
Run / identifier suffix & Query configuration \\
\midrule
R1 / 20260622T224946Z & M1 \\
R2 / 20260622T225354Z & Conversational resolver + M1 + raw \\
R3 / 20260624T022332Z & Focused single-query resolver \\
R4 / 20260625T213110Z & Faithful resolver configuration \\
R5 / 20260626T083357Z & Faithful + M1 + raw (doubled) \\
\bottomrule
\end{tabularx}
\end{table}
The runs share the controller and hybrid retrieval design (Table~\ref{tab:runs}), but later runs also include robustness fixes and use independently generated conversations. Their score differences therefore cannot be attributed solely to query generation. The offline faithful-only condition evaluates query generation under a shared retriever; the deployed runs also include interaction control and \mbox{robustness changes.}

\subsection{What Does Faithful Resolution Change?}
\emph{Motivation.} A standalone query can become less faithful as it becomes more explicit. The offline comparison examines whether preserving the utterance helps ranking, and whether additional queries restore passages missed by the narrower formulation.

\emph{Results.} The faithful query obtains nDCG@10 of 0.231, compared with 0.221 for the combined resolver and 0.171 for the focused resolver (Table~\ref{tab:offline}). Its recall is lower than the combined resolver's. The combined configuration adds M1 and the raw utterance and doubles the contribution of faithful-query passages at the same rank. It increases recall from 0.448 to 0.519, while nDCG@10 decreases from 0.231 to 0.228. The combined resolver retains the highest non-reference recall, 0.581.
\begin{table}[t]
\caption{Offline retrieval results. The combined configuration doubles the contribution of faithful-query passages at the same rank. Bold marks the best non-reference value. We do not report precision because its calculation counted passages judged \mbox{non-relevant as relevant.}}
\label{tab:offline}
\small
\begin{tabularx}{\columnwidth}{@{}Xcc@{}}
\toprule
Query strategy & nDCG@10 & Recall@1,000 \\
\midrule
Manual reference & 0.345 & 0.588 \\
M1 & 0.153 & 0.427 \\
Combined resolver & 0.221 & \textbf{0.581} \\
Focused resolver & 0.171 & 0.457 \\
Faithful query alone & \textbf{0.231} & 0.448 \\
Faithful + M1 + raw (doubled) & 0.228 & 0.519 \\
\bottomrule
\end{tabularx}
\end{table}

\emph{Discussion.} These results show a ranking--recall tradeoff, with no uniform offline winner. Inspection of resolver examples motivated the restriction on added context, but does not measure the frequency or causal contribution of over-elaboration. The P@1 and P@3 calculations counted every passage with a relevance judgment as relevant, including passages judged non-relevant. We therefore do not report \mbox{those precision scores.} The nDCG and recall functions use relevance grades explicitly. We report mean offline scores \mbox{without confidence intervals.}

\subsection{Does the Choice Transfer to Dialogue?}
\emph{Motivation.} We examine whether the retrieval gains observed in our offline experiments also improve response quality during a conversation. We compare retrieval scores and response assessments \mbox{across our submissions.}

\emph{Retrieval results.} R5 obtains the highest mean score in every retrieval column of Table~\ref{tab:official}. Its nDCG@5 is 0.534 against 0.507 for R2, the strongest alternative, a relative improvement of 5.3\%. Its per-type nDCG@5 scores are 0.478, 0.468, 0.482, and 0.709, each the highest among our submissions for that type. Table~\ref{tab:types} reports that consistency by user type. The advantage over R2 varies from 0.012 for u1 to 0.057 for u2; R5 does not dominate \mbox{every metric within every type.}
\begin{table}[t]
\caption{Official nDCG@5 by user type. R5 leads each column among our submissions.}
\label{tab:types}
\small
\begin{tabularx}{\columnwidth}{@{}Xcccc@{}}
\toprule
Run & u1 & u2 & u3 & u4 \\
\midrule
R1 & 0.424 & 0.390 & 0.440 & 0.622 \\
R2 & 0.465 & 0.412 & 0.465 & 0.688 \\
R3 & 0.444 & 0.412 & 0.422 & 0.692 \\
R4 & 0.461 & 0.434 & 0.386 & 0.636 \\
R5 & \textbf{0.478} & \textbf{0.468} & \textbf{0.482} & \textbf{0.709} \\
\bottomrule
\end{tabularx}
\end{table}
\begin{table}[t]
\caption{Official LLM-judged retrieval scores, equally averaged over u1--u4. Bold marks the best of our five submissions. P(2)@5 denotes the organizer's P(rel=2)@5.}
\label{tab:official}
\small
\begin{tabularx}{\columnwidth}{@{}Xccccc@{}}
\toprule
Run & nDCG@5 & nDCG & AP & RR & P(2)@5 \\
\midrule
R1 & 0.469 & 0.200 & 0.102 & 0.728 & 0.527 \\
R2 & 0.507 & 0.237 & 0.127 & 0.727 & 0.592 \\
R3 & 0.493 & 0.221 & 0.113 & 0.718 & 0.561 \\
R4 & 0.479 & 0.221 & 0.112 & 0.715 & 0.549 \\
R5 & \textbf{0.534} & \textbf{0.257} & \textbf{0.136} & \textbf{0.736} & \textbf{0.602} \\
\bottomrule
\end{tabularx}
\end{table}

\emph{Response results.} R5's dialogue personalization score is 3.864 on the five-point scale, while mixed initiative is 1.659. Its overall user satisfaction is 3.614, below R1's 3.773 (Table~\ref{tab:response}). The response-rubric assessment also distinguishes dimensions: R5 scores 4.100 for relevance and usefulness, but 2.852 for engagement. Its overall subtopic quality is 3.583, below R4's 3.613. These dimensions are reported separately because they measure different \mbox{aspects of the interaction.}
\begin{table}[t]
\caption{Official dialogue scores (1--5), mean over 44 assessments per run. Better retrieval in R5 does not coincide with the highest satisfaction.}
\label{tab:response}
\small
\begin{tabularx}{\columnwidth}{@{}Xccc@{}}
\toprule
Run & Personalization & Mixed initiative & Satisfaction \\
\midrule
R1 & 3.841 & \textbf{1.750} & \textbf{3.773} \\
R2 & 3.818 & 1.705 & 3.682 \\
R3 & \textbf{3.909} & 1.727 & 3.591 \\
R4 & 3.864 & 1.568 & 3.636 \\
R5 & 3.864 & 1.659 & 3.614 \\
\bottomrule
\end{tabularx}
\end{table}

\emph{Discussion.} The official retrieval ordering favors weighted fusion, while response scores identify a different limitation. Low mixed-initiative and engagement scores motivate examining the controller, but do not establish that its architecture caused those scores. No human assessment or judge-sensitivity analysis is available to determine whether the small response differences reflect user preferences. The official precision measure has its own relevance criterion and is independent of the excluded offline calculation.

\section{Practical Considerations}
Query generation, passage ranking, personal-fact selection, and interaction control are separate components in our system. The experiments inform choices within these components, while the submitted runs evaluate \mbox{their combined behavior.}

\subsection{Architectural Choices}
\emph{Passage ranking.} Our experiments with pretrained cross-encoders, listwise LLM reranking, dense reranking, and pseudo-relevance feedback did not establish an improvement over the fused ranking. HyDE~\citep{Gao:2023:ACL} improved recall without a corresponding nDCG gain. A query-conditioned nonlinear scorer inspired by Hypencoder~\citep{Killingback:2025:SIGIR} also underperformed its baselines in a five-fold comparison over conversations. These findings informed the decision to retain the BM25 and SPLADE fusion. The experiments used different setups, so they do not establish a general ranking of these approaches. An in-domain cross-encoder \mbox{remains untested.}

\emph{Personal-fact selection.} We compared generate-then-verify selection and an LLM relevance score with a calibrated threshold against a conservative binary selector. The best calibrated F1 was 0.38, compared with 0.46 for binary selection. We therefore retained the binary selector for facts extracted from the conversation. The decider selects provided PTKB statements separately, so this comparison evaluates one part of the memory component. Its effect on the submitted runs' personalization scores \mbox{was not measured separately.}

\emph{Search refinement.} The controller can request a refined search, but the configured query generator uses the turn context and takes precedence over the controller's proposed query strings. A refinement request therefore does not ensure that the next query incorporates the controller's suggested changes. This limits how the controller can adapt retrieval to the evidence already returned. We did not compare alternative refinement mechanisms, so its effect on mixed-initiative scores \mbox{remains an open question.}

\subsection{Interpreting the Results}
The offline experiments compare query formulations under a shared retriever. They support the observed tradeoff between ranking quality and recall on iKAT-2023. The combined configuration changes both the query set and the relative contribution of the faithful query, so the comparison does not establish how much each change contributes to the result. A comparison that varies the contribution while keeping the query set fixed would \mbox{separate these effects.}

The interactive scores describe the complete submitted configurations, including interaction control and later robustness changes. Better retrieval in R5 did not coincide with the highest satisfaction score. This finding motivates evaluating retrieval and response quality separately; it does not identify which component caused the response differences. The evaluation uses simulated users and LLM-based assessments. Human preferences and agreement with human relevance judgments \mbox{were not evaluated.}

\begin{samepage}
\section{Conclusion}
Faithful query resolution preserves the current information need, while weighted fusion adds retrieval breadth through separate query formulations. In our iKAT SCAI 2026 submissions, the weighted configuration obtains mean official nDCG@5 of 0.534, exceeding the strongest alternative by 5.3\%. Our offline experiments show the cost of that breadth as a slight nDCG@10 reduction relative to the faithful query alone. The response assessments do not show a corresponding satisfaction advantage. Query fidelity and retrieval breadth can be controlled separately, while their effects on conversational utility require separate assessment.
\par
\end{samepage}
\bibliographystyle{ACM-Reference-Format}
\bibliography{references}
\end{document}